\documentclass[12pt]{article}
\usepackage{amsfonts}
\usepackage{amssymb}
\usepackage{epsfig}

\newcommand{\beq}{\begin{equation}}
\newcommand{\eeq}{\end{equation}}
\newcommand{\beqn}{\begin{eqnarray}}
\newcommand{\eeqn}{\end{eqnarray}}
\newcommand{\bea}[1]{\beq\begin{array}{#1}}
\newcommand{\eea}{\end{array}\eeq}

\newcommand{\sign}{\mathop{\rm sign}}

\newcommand{\tr}{\mathop{\rm Tr}}

\newcommand{\NP}[3]{{\it Nucl. Phys. }{\bf #1} (#2) #3}
\newcommand{\NPPS}[3]{{\it Nucl. Phys. Proc. Suppl. }{\bf #1} (#2) #3}
\newcommand{\PL}[3]{{\it Phys. Lett. }{\bf #1} (#2) #3}
\newcommand{\PRL}[3]{{\it Phys. Rev. Lett. }{\bf #1} (#2) #3}

\newcommand{\PR}[3]{{\it Phys. Rev. }{\bf #1} (#2) #3}
\newcommand{\JL}[3]{{\it JETP Lett. }{\bf #1} (#2) #3}

\newcommand{\JHEP}[3]{{\it JHEP }{\bf #1} (#2) #3}

\begin{document}
\date{}
\title{Fine tuned vortices in lattice $SU(2)$ gluodynamics
\vskip-40mm
\rightline{\small ITEP-LAT/2002-26}
\rightline{\small MPI-PhT/2002-76}
\vskip 30mm
}

\author{F.V.~Gubarev, A.V.~Kovalenko, M.I.~Polikarpov, S.N.~Syritsyn \\
{\small Institute of Theoretical and  Experimental Physics,}\\
{\small B.~Cheremushkinskaja 25, Moscow, 117259, Russia}\\
\\
V.I.~Zakharov \\
{\small Max-Planck Institut f\"ur Physik,}\\
{\small F\"ohringer Ring 6, 80805, M\"unchen, Germany}
}

\maketitle
\begin{abstract}\noindent
We report measurements of the action associated with center vortices in
$SU(2)$ pure lattice gauge theory. In the lattice units the excess of the
action on the plaquettes belonging to the vortex is approximately a constant,
independent on the lattice spacing $a$. Therefore the action of the center
vortex  is of order $A/a^2$, where $A$ is its area. Since the area $A$ is known
to scale in the physical units, the measurements imply that the suppression
due to the surface action is balanced, or fine tuned to the entropy factor
which is to be an exponential of $A/a^2$.

~

\noindent
PACS numbers: 11.15.-q, 11.15.Ha, 12.38.Gc
\end{abstract}

\section{Monopoles and vortices}

Lattice measurements allow for a direct study of field fluctuations in the
vacuum state of Yang-Mills theories. Generically, the probability
to find a field configuration is a product of the entropy and action factors:
\beq
\label{probability}
P~=~ \exp({\mathfrak S}) \,\cdot\,  \exp(-S)\,,
\eeq
where $S$ is the action and the entropy $\exp({\mathfrak S})$ is the number
of ways in which the field configuration can be realized.
Let us mention two simple examples to illustrate (\ref{probability}).
In case of instantons, the (classical) action is $S_{inst}=8\pi^2/g^2$ while
the entropy is provided by counting small (quantum) fluctuations in
the instanton background.
Another  example is the zero-point vacuum fluctuations.
They are dominated by the phase space, or by the entropy.

These two examples demonstrate relevance to various fluctuations of two
distinct scales, that is of the lattice spacing $a$ and $\Lambda_{QCD}^{-1}$.
The latter is in physical units and does not depend on the lattice. The lattice
spacing serves as an ultraviolet cutoff. In particular the average action
density $\langle s \rangle$ is ultraviolet divergent,
\beq
\langle s \rangle
~\sim ~ (N_c^2 - 1) \, a^{-4}\,,
\eeq
where $N_c$ is the number  of
colors\footnote{
{}From now on we will consider only the $SU(2)$ case, $N_c = 2$.
}. The ultraviolet divergence, $a^{-4}$, is due to the zero-point fluctuations
and is well known in field theory.
On the other hand, the quasi-classical fluctuations, like instantons, are
driven to the infrared scale of order $\Lambda_{QCD}^{-1}$.

On the lattice, there were also observed fluctuations which are defined as
geometrical objects, that is closed worldlines and closed surfaces. We mean now
monopoles and  center vortices, for review see, e.g., Ref.~\cite{reviews}. What
is common for the monopoles and vortices is that they are defined not within
the original theory but within a projected theory. In case of monopoles one
projects the original $SU(2)$ to $U(1)$ theory and in case of the P-vortices --
to $Z(2)$ gauge theory. In more detail, the monopoles and P-vortices are
defined as follows\footnote{
Throughout this paper we consider  only the case of so called
indirect maximal center gauge, see, e.g., Ref.~\cite{IMC}.
}. First, one fixes the maximal Abelian gauge by maximizing the functional:
\beq
\label{mag}
R_{Abel} = \sum\limits_{x,\mu} \tr[\, U_{x,\mu} \sigma^3 U^+_{x,\mu} \sigma^3 \,]\,,
\eeq
where $U_{x,\mu}$ are link matrices. Next, Abelian projection is made by replacing
\beq
\label{mag-projection}
U_{x,\mu} \, \rightarrow \, U^{Abel}_{x,\mu} \,=\, \zeta_{x,\mu}\,/\,|\zeta_{x,\mu}|\,,
\qquad
\zeta_{x,\mu} \,=\, \tr[\,(1+\sigma^3)\,U_{x,\mu}\,]\,.
\eeq
Then the monopoles are defined as
the topological defects in $U^{Abel}_{x,\mu}$ fields. By construction they
form a closed world-lines on the dual lattice.

Since the functional (\ref{mag}) leaves the $U(1)$ gauge freedom unfixed one
can fix the gauge further by maximizing
\beq \label{center}
R_{center} \,=\, \sum\limits_{x,\mu} \, (\Re e~ U^{Abel}_{x,\mu})^2
\eeq
with respect to $U(1)$ gauge rotations. Then the maximal center
projection amounts to replacing
\beq
\label{center-projection} U^{Abel}_{x,\mu} \rightarrow Z_{x,\mu} = \sign~\Re
e~U^{Abel}_{x,\mu}\,.
\eeq
Plaquettes constructed in a standard way from
$Z_{x,\mu}$ have values $\pm 1$. Finally, P-vortices are defined as union of
all the negative plaquettes and are closed surfaces on the dual lattice.

Knowing only definitions of these geometrical objects, lines and surfaces, it
is not easy to figure out what kind of physics can be revealed  by their
studies. However, there emerged phenomenology indicating that there are some
physical objects detected through the projections. ``Physical'' in the present
context means first of all that the area, (respectively, length) of the
percolating vortices (monopoles) is in physical units, see,
e.g.,~\cite{bornyakov,langfeld}:
\beq \label{scale}
A_{vort}~=~6\,\rho_{vort}\cdot V_4\,,
\qquad
\rho_{vort}~\approx~4~(fm)^{-2}\,,
\eeq
\beq
\label{scale-mono}
L_{mon}~=~4\,\rho_{mon}\cdot V_4\,,
\qquad
\rho_{mon}~\approx~6~(fm)^{-3}\,,
\eeq
where $V_4= a^4 L^4$ is the volume of the lattice.

Note that at this stage we do not have yet any information on the
action and entropy factors entering Eq.~(\ref{probability}) and
one is free to speculate theoretically about them.
The common viewpoint is that there are objects of the size
of order $\Lambda_{QCD}^{-1}$ behind mathematically thin lines \cite{action}
or vortices \cite{faber} defined in the
projected theories. The thin geometrical objects then mark these
bulky structures and their position within the ``thick''
fluctuations is more or less accidental. The only indirect evidence in favor of
the physical objects being thin is that both monopoles and vortices
generate a linear piece in the heavy quark potential even at short distances~\cite{gpz}.

On the other hand, one can try to measure the action and even the entropy
associated with the geometrical objects directly.
Such measurements have been performed mostly for the monopoles, see
\cite{anatomy,boyko}, and the results can be interpreted
only as fine tuning \cite{vz}.
Namely, both the action and the entropy are ultraviolet divergent\footnote{\label{endnote22}
A caveat here is that we are interpreting the measurements on the presently available
lattices only and consider appearance of negative powers of the lattice spacing $a$
as a sign of ultraviolet divergence. The actual limit $a\to 0$ can be different if
the observed pattern changes at smaller $a$.
} but cancel each other to order $\Lambda_{QCD}$:
\beq \label{tuning}
|{\mathfrak S}_{mon} - S_{mon}| = |({\mathfrak s}_{mon} - s_{mon})| \cdot L/a
\sim  \Lambda_{QCD} \cdot L\,,
\eeq
where $L$ is the length of the monopole
trajectory. Moreover the action associated with the monopoles is measured on
the lattice (see Ref.~\cite{anatomy} for details) and turns
out to be ultraviolet divergent. The entropy factor is then easy to calculate,
${\mathfrak s}_{mon}=\ln 7$ and this factor corresponds to the number of trajectories
of the same length $L$.

In this letter we extend the exploration of the anatomy of the
geometrically defined fluctuations to the case of vortices.
Namely, we study the action density both on the plaquettes
belonging to the vortex and on the adjacent plaquettes as a
function of the lattice spacing $a$.
Our main result is that the  excess of the action density on the vortex is
independent on the lattice spacing if expressed in the lattice
units:
\beq
\label{excess}
\langle S_{vort}\rangle - \langle S_{vac}\rangle =
0.540 \pm 0.004 \qquad \mathrm{[lattice~units]}\,,
\eeq
where $\langle S_{vort}\rangle$ is the average value of the non-Abelian action density
on the plaquettes belonging to the vortex and $\langle S_{vac}\rangle$
is the plaquette action averaged over the whole lattice. Note that the action excess
for a particular value of $\beta = 2.4$ was first measured in Refs.~\cite{correlation,giedt}
and we agree with these results. Our main new point is the measurement of the action
of the vortices as function of the lattice spacing $a$.
We also calculate the excess of the action on the plaquettes which are
nearest to the P-vortex world-sheet. It turns out that this quantity is more or
less consistent with zero, see next Section.

\section{Numerical Results}
\label{section2}

We have performed our calculations in pure $SU(2)$ lattice gauge
theory for $2.35\le \beta \le 2.6$.
The lattice spacing $a$ is fixed using the standard values \cite{sigmal} of the
lattice string tension, which in physical units is $\sqrt{\sigma} = 440\,\mathrm{MeV}$.
At each value of $\beta$ we have considered 20 statistically independent
configurations generated on symmetric $L^4$ lattices.
The lattice size was $L=16$ at $\beta = 2.35$,
$L=24$ for $\beta = 2.4$, $2.45$, $2.5$ and $L=28$ at $\beta = 2.55$, $2.6$.
The indirect maximal center gauge~\cite{IMC}
was employed, and the definition of the gauge is given above.

To fix the maximal Abelian and the maximal center gauges we have used the
simulated annealing algorithm~\cite{SA}. For maximal Abelian gauge 20 gauge
copies of each $SU(2)$ field configuration were considered  and the simulated
annealing algorithm was applied to each copy. For $U(1)$ gauge fixing only the
configuration which corresponds to the maximal value of the functional
(\ref{mag}) was considered. Furthermore, only one gauge copy of the Abelian
configuration was taken into account for fixing maximal center gauge, since we
checked that P-vortex density varies by less than $1\%$ for various $U(1)$
gauge copies.

First, we discuss the P-vortex density,
$\rho=\langle N_{PV} / (6L^4 a^2) \rangle$,
where $N_{PV}$ is the number of plaquettes occupied by P-vortices.
The dependence of $\rho$ on the lattice spacing is shown on the Fig.~1.
Note that all quantities are in physical units.
It is clearly seen that $\rho$ tends to the limit (\ref{scale}) as $a \to 0$.

\begin{figure}[t]
\centerline{\epsfig{file=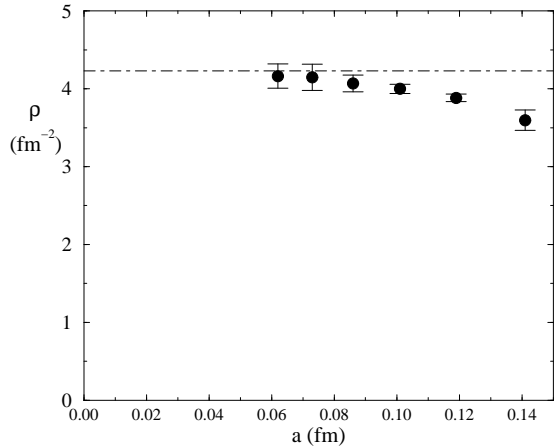,width=0.45\textwidth}}
\caption{The density of P-vortices vs. lattice spacing.}
\end{figure}

Next we consider the average action density, $S_{PV}$, on the plaquettes dual to those
forming P-vortices (we refer to these plaquettes as 'P-vortex plaquettes' below).
It occurs that it is much larger then the average plaquette
action, $S_{vac}=\beta(1-\langle \tr U_P\rangle/2)$. The dependence of the difference
$S_{PV}-S_{vac}$ on the lattice spacing is shown on Fig.~2 by circles.

In order to probe the internal structure of the vortices we measured the
average action density near P-vortex world-sheet. In more details, we have studied two
types of the nearest plaquettes: the first type, 'side plaquettes', lie in the
same plane as the P-vortex plaquette and have a common link with it; the second
type, 'closest plaquettes', have a common link with the P-vortex plaquette, but
are perpendicular to it. The two types of the plaquettes are depicted in
Fig.~3. The corresponding excess of the action is shown on Fig.~2 by the up and
down triangles.

Moreover, as first observed in \cite{correlation,giedt}, the vortices and
monopoles are strongly correlated with each other for $\beta = 2.4$. We confirm
the strong correlation of the monopoles and vortices for other values of
$\beta$. Moreover, we measure the fraction of $S_{PV}$ which is due to the
monopoles. We define $S^{mon}_{PV}$ as the average action density on the P-vortex
plaquettes which have a common link with a monopole trajectory. It turns out
that the quantity $S^{mon}_{PV} - S_{vac}$ (shown on Fig.~2 by squares) is even
larger than $S_{PV} - S_{vac}$ implying that indeed a large fraction of the
vortex action is due to the Abelian monopoles. If we exclude the P-vortex
plaquettes which touch the monopole trajectory, the corresponding average
action is lower than $S_{PV}$ (diamonds on the Fig.~2).

\begin{figure}[t]
  \begin{tabular}{p{0.5\textwidth} p{0.5\textwidth}}
    \begin{center}
     \epsfig{file=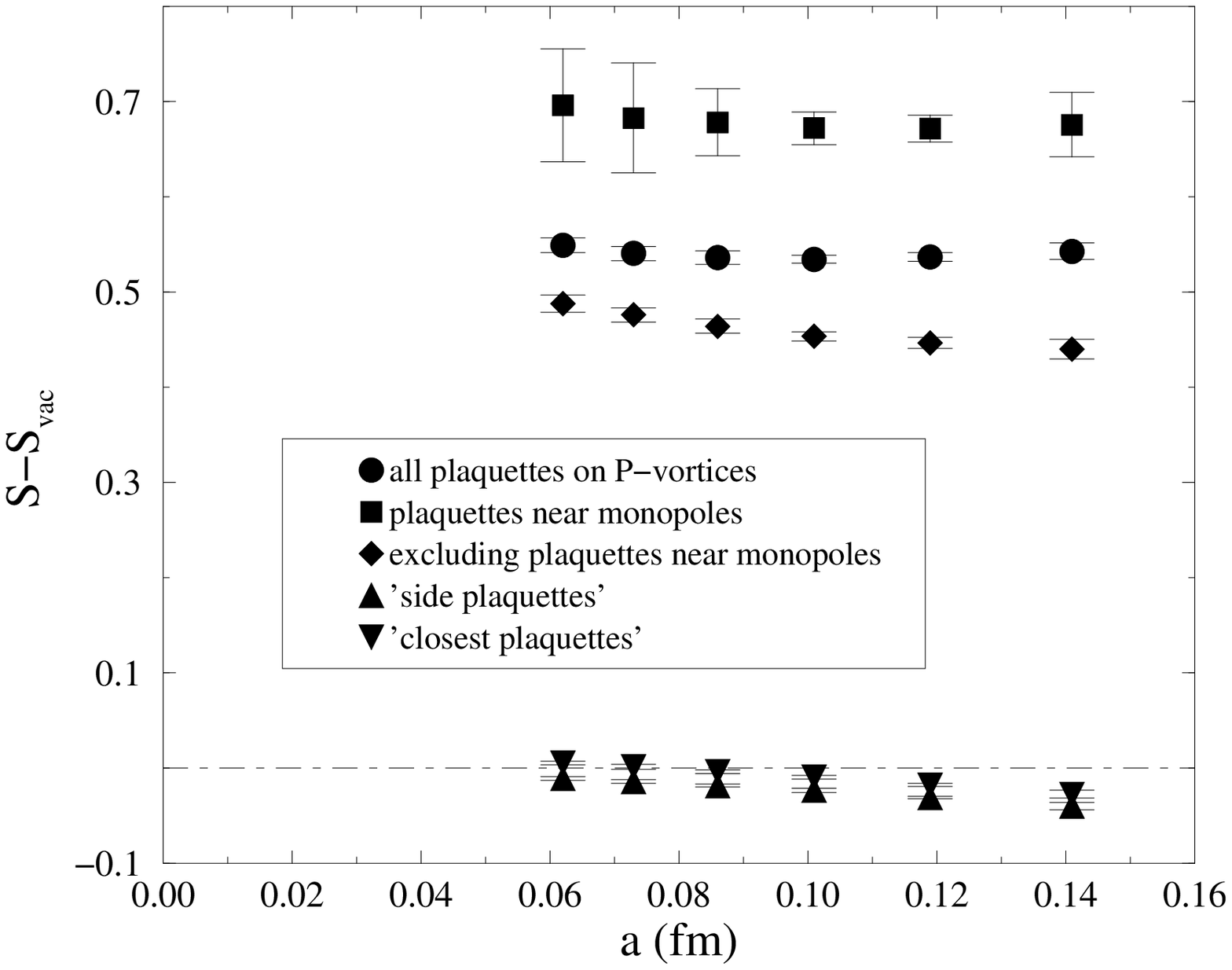,width=0.5\textwidth}
     Figure 2: The excess of the action on and around P-vortices.
    \end{center}
    &
    \begin{center}
     \epsfig{file=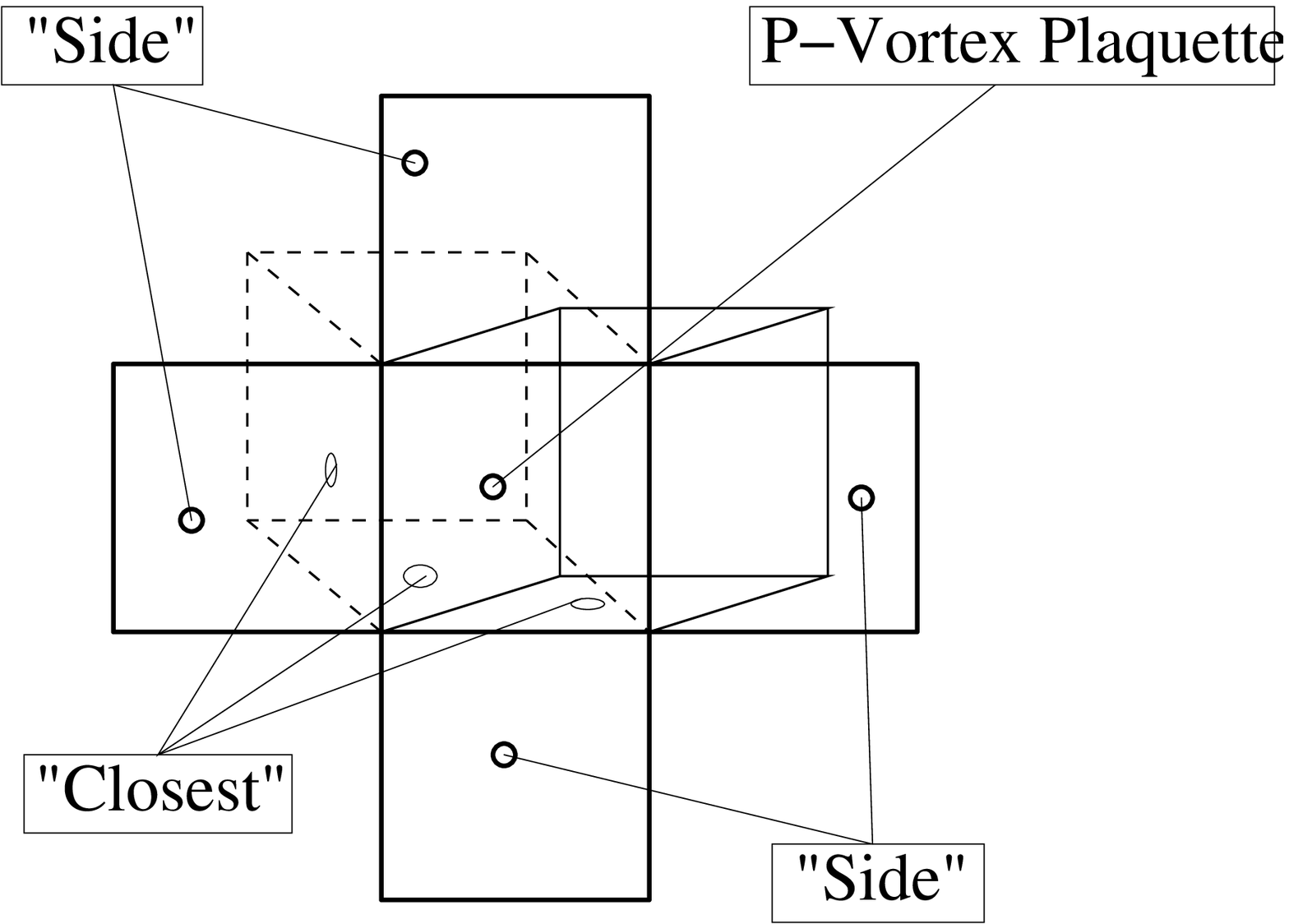,width=0.5\textwidth}
     Figure 3: Two types of the nearest to P-vortex plaquettes (see the text).
    \end{center}
  \end{tabular}
\end{figure}


\section{Discussions}

We see that at presently available lattices the vortices appear as infinitely
thin objects with no sign of any internal structure\footnote{
See, however, footnote (\ref{endnote22}) for a reservation.
}. Our measurements allow to conclude that the vortex thickness is
\beq \label{thickness}
R_{vort} ~\lesssim~ 0.06\mathrm{~fm}\,,
\eeq
where $0.06\mathrm{~fm}$ is the smallest lattice
spacing used in our simulations. Note that a similar estimate of the monopole
size was obtained in \cite{anatomy}.

Taken at face value, the lattice data imply that we are dealing with
infinitely thin (and in this sense "fundamental") strings which populate the vacuum.
Assuming that at the ultraviolet scale the surfaces can be considered
independent of the rest of the vacuum, the ultraviolet divergence of the action
is to be canceled by a corresponding entropy factor:
\beq
|{\mathfrak S}_{vort} - S_{vort}| = |{\mathfrak s}_{vort} - s_{vort}| \cdot A/a^2
\sim \Lambda^2_{QCD} \cdot A\,,
\eeq
similar to the case of the monopoles~\cite{vz}.

It is worth emphasizing that we define thickness of the vortex in terms
of the distribution of the non-Abelian action. One can define the vortex thickness
in terms of the flux carried by the vortex. Then the vortex seems not to
be localized to the cutoff scale. The corresponding discussion can be found
in Refs.~\cite{langfeld,IMC}.

\mbox{}From the theoretical point of view,  interpretation of the results
obtained represents a challenge. Indeed, if one introduces random surfaces on
the lattice with action proportional to the area they appear unstable with
respect to the decay into branched polymers (see, e.g., Ref.~\cite{ambjorn} for
review). In other words, the model of random surfaces collapses in fact to the
theory of single non-interacting scalar particle. Note that this remark applies
to the random surfaces with limited genus. The genus of the percolating
P-vortices, on the other hand, grows with the lattice volume. However, this
growth is associated with the distances of the order $\Lambda_{QCD}^{-1}$
\cite{genus} while the instability mentioned above develops at the ultraviolet
scale, that is at the scale of the lattice spacing $a$.

The nearest extension of the bosonic string is the inclusion of an extrinsic
curvature term~\cite{curvature}. Namely, adding the curvature term one can get
a fine tuned surfaces. Moreover,  such a model is successful as a
phenomenological statistical description of strings in four
dimensions\footnote{ 
Consideration of bosonic strings with extrinsic curvature as a model for P-vortices
can be found in Ref.~\cite{P-curvature}.
} \cite{curvature-lattice}.

A key to understanding the structure of the thin vortices
could be provided by the observation of the strong correlation between
the vortices and monopoles (which are originally defined as
independent geometrical objects).
In particular, basing on the fact that the monopole-associated plaquettes
are  "hotter" than the average (see Sect.~\ref{section2})
one is tempted to assume that the monopoles are associated with
the `creases' of the P-vortex
world-sheet and correspond to the extrinsic curvature term in the P-vortex action.

To summarize, we have observed surfaces whose thickness is smaller than the
presently available resolution, $a$ and whose area scales in the physical
units. Moreover, the thickness is defined in terms of the original non-Abelian
action. The coexistence of the two scales, that is $a$ and $\Lambda_{QCD}$ can
be called fine tuning. A remarkable feature of the surfaces is that they are
associated also with the monopole trajectories. In turn the monopoles condense
and in this sense correspond to the tachyonic mode in the field-theoretical
language. Therefore, we can say that there are indications that in case of the
four dimensional gluodynamics the tachyonic mode is confined to a two
dimensional surface.

\section{Acknowledgments}
The authors are grateful to V.G.~Bornyakov, G.~Marchesini and E.T.~Tomboulis
for useful discussions. This work is partially supported by grants RFBR
02-02-17308, RFBR 01-02-117456, RFBR 00-15-96-786, INTAS-00-00111, and CRDF
award RPI-2364-MO-02. A.V.K. and S.N.S. are partially supported by CRDF award
MO-011-0.


\end{document}